\documentclass[preprint,journal,pdftex]{vgtc}       

\usepackage{flushend}
\usepackage{amsmath}
\usepackage{amssymb}
\usepackage{color}
\usepackage{mathptmx,gensymb,mathrsfs}
\usepackage{graphicx}
\usepackage{subfigure}
\usepackage{float}
\usepackage{times}
\usepackage{listings}
\usepackage{url}
\usepackage[T1]{fontenc}
\usepackage[utf8]{inputenc}

\definecolor{redcol}{rgb}{.7,0,0}

\definecolor{LightGray}{gray}{.90}

\newfloat{floatbox}{ftbh}{lob}
\floatname{floatbox}{Box}

\title{Interactive Visualization and Simulation of Astronomical Nebulae}

\author{Stephan Wenger, Marco Ament, Wolfgang Steffen, Nico Koning, Daniel Weiskopf, and Marcus Magnor}

\authorfooter{
\item S. Wenger and M. Magnor are with the Computer Graphics Lab, TU Braunschweig, Germany. \\ E-mail: \{wenger, magnor\}@cg.cs.tu-bs.de.
\item M. Ament and D. Weiskopf are with Visualization Research Center (VISUS), University of Stuttgart, Germany. \\ E-mail: \{Marco.Ament, Daniel.Weiskopf\}@visus.uni-stuttgart.de.
\item W. Steffen is with the the Instituto de Astronom\'{\i}a, Universidad Nacional Aut\'{o}noma de M\'{e}xico, Ensenada, B.C., Mexico. \\ E-mail: wsteffen@astrosen.unam.mx.

Preprint. Published in \emph{Computing in Science \& Engineering}, vol.~14, no.~3, pp.~78--87, May~2012.
Copyright 2012 IEEE.

\url{http://dx.doi.org/10.1109/MCSE.2012.52}
}

\abstract{
Interactive visualization and simulation of astrophysical phenomena
enable digital planetariums and television documentaries
to take their spectators on a journey into deep space
and explore the astronomical wonders of our universe in 3D.
}

\begin{document}

\maketitle

When thinking about astronomical objects, the first thing that comes to mind is probably stars.
Interstellar space, however, is also full of other fascinating phenomena:
among the most popular ones are nova and supernova remnants as well as emission and reflection nebulae.
What makes these objects so attractive is their intriguingly complex, intricate, and often colorful structure.
Observing and investigating them is not only an aesthetic pleasure
but helps physicists and astronomers deduct information about our universe and its laws of nature,
for example by exploring cosmological effects from the theory of special~\cite{mueller2011special} and
general~\cite{mueller2011general} relativity.

With the advent of the Hubble Space Telescope (HST),
high-quality imagery of many distant objects has become available and popular.
Unhindered by the Earth's atmosphere,
the HST is able to capture images of unprecedented resolution.
Through the use of filters, objects can be observed at distinct wavelengths,
such as the emission lines of various ions outlining the distribution of different elements
in the gaseous cloud of astronomical nebulae.
Combining multiple such photographs produces the colorful, yet typically falsely-colored images
we know from NASA and HST press releases (\url{http://hubblesite.org/}).
Not only space-based telescopes yield spectacular new images,
but also adaptive optics reducing the blurring effect of the atmosphere
have led to high resolution images in the optical and near-infrared region of the spectrum.
With highly sensitive CCD cameras, even amateur astronomers with relatively small telescopes
are today taking incredible images.

One fundamental limitation of images captured from within our solar system
is that they are inherently flat and two-dimensional.
Because of the huge distance to most astronomical objects of interest,
our fixed vantage point provides no useful parallax:
wherever and whenever we take a photo of an interstellar object, its view remains the same.
For nebulae, galaxies, and other complex objects,
this makes it notoriously hard to identify their three-dimensional structure.

Three-dimensional structural information, however, is the key to many astrophysical questions.
Scientists use 3D models of galaxies and nebulae to validate their theories about
the formation and evolution of these objects.
In addition, the wide public appreciates high-quality 3D visualizations of astronomical phenomena.
Television documentaries disseminating advances in physics and space sciences
make frequent use of animations to explain our universe.
Modern planetariums have evolved from ``indoor starry skies''
to full-blown surround-video dome theaters
with high-resolution digital projection systems and cinema sound systems~\cite{magnor2010planetariums}.
Many planetariums already show more full-dome movie productions than traditional astronomy shows.
Such educational shows can be spiced up with astronomical 3D content.
Whether the show is pre-rendered or interactively presented,
three-dimensional visualizations of otherwise abstract and complex subjects
stimulate imagination and aid in the process of conveying scientific information.

Artists and astronomers face several challenges when creating three-dimensional 
visualizations of astronomical content.
Sometimes it is sufficient for simulated data to provide an impression
of what a general class of objects looks like and how it evolves in principle;
often, however, it is more interesting to show the reconstruction of an actual astronomical object.
In some cases, it is enough to present plausible but not scientifically exact reconstructions of existing objects;
in others, scientific accuracy is called for.
For an audience that is used to high-resolution renderings and high-quality special effects,
the realism and resolution of the visualization is of utmost importance,
often driving current hardware and algorithms to their limits.
Especially for interactive presentations,
rendering algorithms do not only have to model the physical light propagation and
image formation as accurately as possible,
but the algorithms also need to do so at real-time framerates for images the size of several megapixels.

While these challenges can be diminished with more or faster hardware,
the fundamental problem in astronomical modeling and visualization
is that of missing 3D information about astronomical objects.
In past and current research, several methods have been proposed
for generating three-dimensional visualizations from two-dimensional imagery in a more or less automatic way.
The scientific accuracy of the results can, however, only be judged by human specialists,
and many complex objects can only be modeled by hand.
Specialized modeling tools aid astronomers and artists in this task.
Other interactive tools support the simulation of fictitious objects to illustrate general concepts.
Last but not least, sophisticated rendering algorithms process the resulting models
to generate aesthetically pleasing and scientifically sound visualizations on the screen in real-time.

\begin{floatbox}
\centering\noindent\colorbox{LightGray}{\begin{minipage}{0.98\linewidth}
\label{box:doppler}
\subsection*{DOPPLER EFFECT}

The Doppler effect plays a key role in the observation of the universe.
It describes the change of frequency and wavelength
for radiation from an object that moves towards or away from the observer.
Without the Doppler effect, we would probably be unaware of the \emph{Big Bang}.
When the relative speed along the line between a source of radiation and the observer is $v$,
the observed frequency of a spectral line of frequency $f$ changes by
\begin{equation*}
\Delta f = f \cdot \frac{v}{c} \,.
\end{equation*}
Here, $\Delta f$ is the change in observed frequency and $c$ is the speed of light.
By measuring $\Delta f$ and solving for $v$,
the Doppler effect can be used to measure relative speed along the line of sight.

For some astronomical objects, the speed towards or away from the observer
correlates with the distance from the observer.
For instance, in ballistic explosions the velocity $\mathbf{v}$ may be different for each particle
but is constant over time for every part of the expanding material.
Therefore, the position $\mathbf{p}$ of a particle varies with the time $t$ as
\begin{equation*}
\mathbf{p} = \mathbf{v} \cdot t \,.
\end{equation*}

In a ballistic expansion from a single point, the structure of the overall object remains constant;
it is said that the expansion is \emph{homologous}.
Since the time since the explosion is usually unknown,
the velocities do not translate directly into distances from the center of the explosion.
Instead, an additional constraint---typically a symmetry assumption---is needed to fix the factor of conversion.
Alternatively, the expansion velocity in the image plane can be measured directly by comparing two images taken several years apart.
The time of the explosion can then be computed by extrapolating the expansion back to the center of the object.

\end{minipage}}
\end{floatbox}

\begin{floatbox}
\centering\noindent\colorbox{LightGray}{\begin{minipage}{0.98\linewidth}
\label{box:hydrodynamics}
\subsection*{EULER EQUATIONS}

In computer animation, hydrodynamics is used to simulate various phenomena
such as water waves, swirling clouds in air and water, or fire and fiery explosions.
Except for explosions, all these fluid effects can be convincingly simulated
using the \emph{incompressible} Navier-Stokes equations.
A key difference in astrophysical fluid dynamics is that most observable phenomena precisely arise
from the compressibility of the gas in the extreme conditions of interstellar space.
Therefore, the numerical methods commonly applied in computer graphics are unsuitable to simulate astrophysical fluids.
As long as we can ignore magnetic and gravitational forces, the inviscid Euler equations are employed instead, which permit compression:
\begin{equation*}
\frac{\partial \mathbf{U}}{\partial t}
+ \frac{\partial \mathbf{F}}{\partial x}
+ \frac{\partial \mathbf{G}}{\partial y}
+ \frac{\partial \mathbf{H}}{\partial z}
= \mathbf{S}
\end{equation*}
Here $\mathbf{F}$, $\mathbf{G}$, and $\mathbf{H}$ are the fluxes along the cartesian coordinate directions $x$, $y$, and $z$, respectively.
They represent the mass, momentum, and energy that flows between the volume cells per unit time and area.
Furthermore, $\mathbf{U}$ is the total internal energy, $\mathbf{S}$ represent sources and sinks of energy and $t$ denotes time.
The source and sink terms describe heating, cooling, and other possible energy changes
due to external forces or nuclear or chemical reactions.

When appropriate solvers are used, the Euler equations allow for the formation of \emph{shock fronts}.
They are created when matter moves through a medium faster than the speed of sound,
compressing the surrounding gas to many times its original density, heating it and thereby making it visible.
Such shock fronts are ubiquitously observed in space as supernova remnants, jets or stellar winds interacting with the interstellar gas,
producing colorful ``spacescapes" that have become widely known through images from the Hubble Space Telescope.
The software \emph{Shape} incorporates ``astrophysics grade'' hydrodynamics in an interactive 3D modeling and animation framework following the scheme by Raga et al.~\cite{raga2000adaptive} on a uniform cartesian grid.

\end{minipage}}
\end{floatbox}

\section*{3D Reconstruction}
\label{sec:reconstruction}
Astronomers, planetarium show presenters, and directors of documentaries
all benefit from being able to generate 3D visualizations for astronomical phenomena
with the least possible amount of manual work.
The easiest way to create three-dimensional content
is to download a high-resolution image of the desired object from an online database
and feed it into an automatic reconstruction algorithm.
However, the reconstruction algorithm faces a difficult problem:
the image contains only information in two dimensions,
but does not directly provide any insight into depth or distance.
Additional information, in form of a general model of what we expect the object to look like,
is necessary to resolve this ambiguity.
This includes both our knowledge about the emission and transport of light in astronomical objects
and assumptions about their presumed geometry.

How exactly the two-dimensional image is formed depends on many physical properties of the phenomenon.
Most objects of interest are at least partially transparent,
so that the intensity we observe is a superposition of the object brightness at different depths.
Some objects, like reflection nebulae, contain large amounts of interstellar dust
that absorbs or scatters light from surrounding stars.
When the dust is very dense, we can only see the outmost layers of the object before our view is effectively blocked,
and no information about anything behind these layers can be reliably deducted from the image.

On the other hand, many objects contain only small amounts of opaque matter.
Planetary nebulae and supernova remnants, for example, develop from stars close to the end of their lifetime.
The star ejects gas into the surrounding space, sweeping away any potential interstellar dust in its proximity.
Due to the radiation from the central star, parts of the gas become ionized.
When the electrons and ions finally recombine,
light is emitted at characteristic wavelengths depending on the specific kind of ion involved.
Since the gas is highly transparent,
radiation from all depths within the nebula can be observed simultaneously,
and the image contains information about all parts of the object.

\begin{figure*}[ftbh]\centering
\includegraphics[width=.32\linewidth]{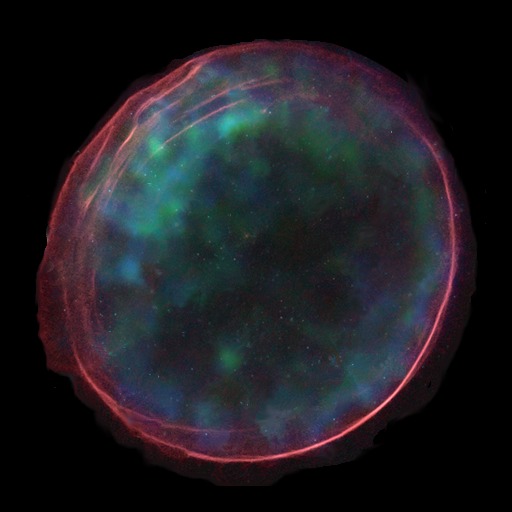}
\includegraphics[width=.32\linewidth]{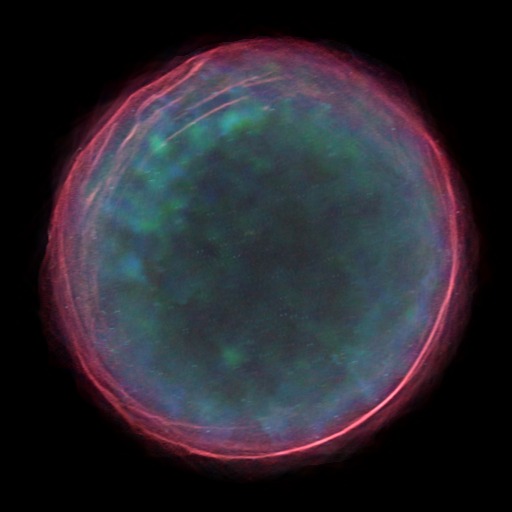}
\includegraphics[width=.32\linewidth]{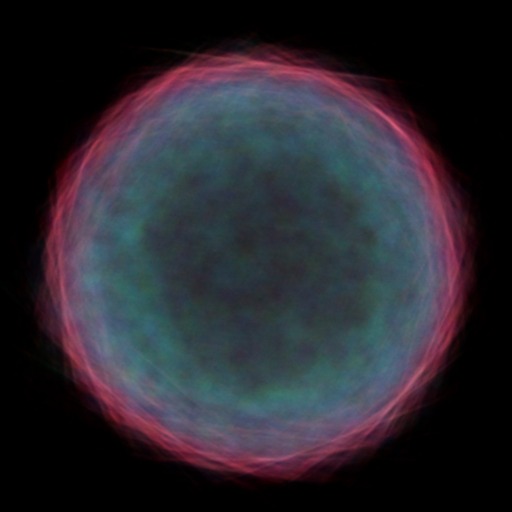}
\caption{Supernova remnant 0509-67.5 in the Large Magellanic Cloud.
From left to right: observed image and reconstruction using spherical symmetry assumption seen from Earth and from space, respectively.
\emph{Original image: NASA, ESA, CXC, SAO, the Hubble Heritage Team (STScI/AURA), and J. Hughes (Rutgers University)}}
\label{fig:eurovis-spherical}\end{figure*}

\begin{figure*}[ftbh]\centering
\subfigure[NGC~6826.]{
\includegraphics[width=.32\linewidth]{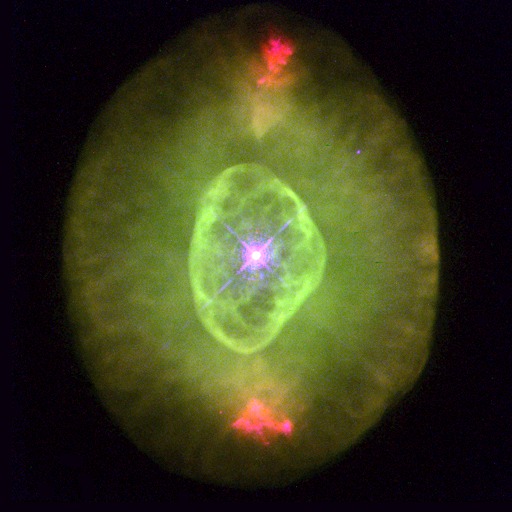}
\includegraphics[width=.32\linewidth]{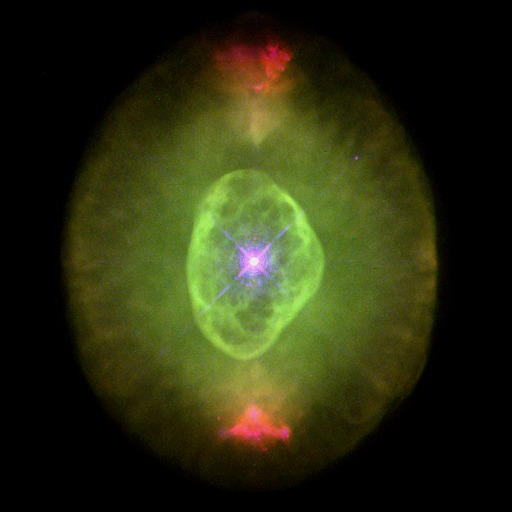}
\includegraphics[width=.32\linewidth]{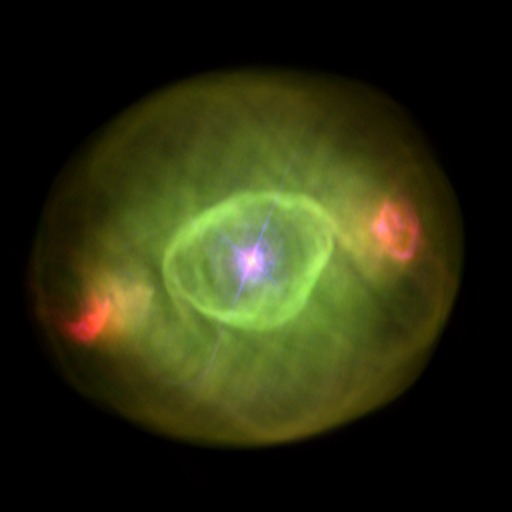}
\label{fig:ngc6826}
}
\subfigure[IC~4406.]{
\includegraphics[width=.32\linewidth]{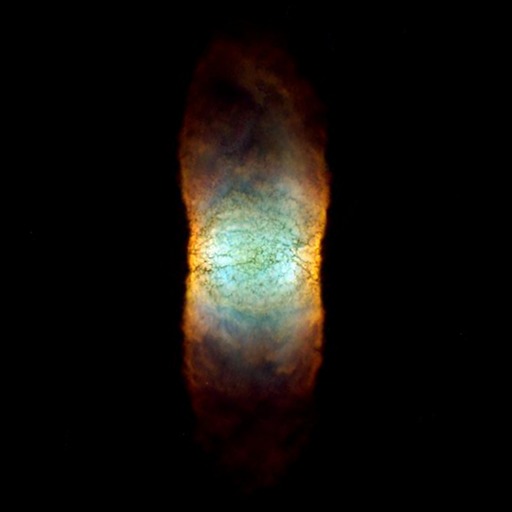}
\includegraphics[width=.32\linewidth]{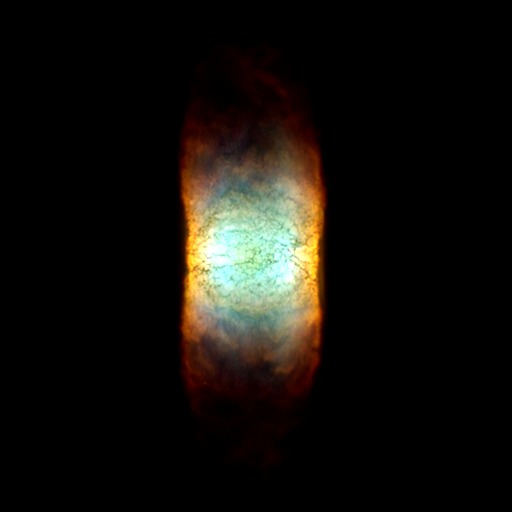}
\includegraphics[width=.32\linewidth]{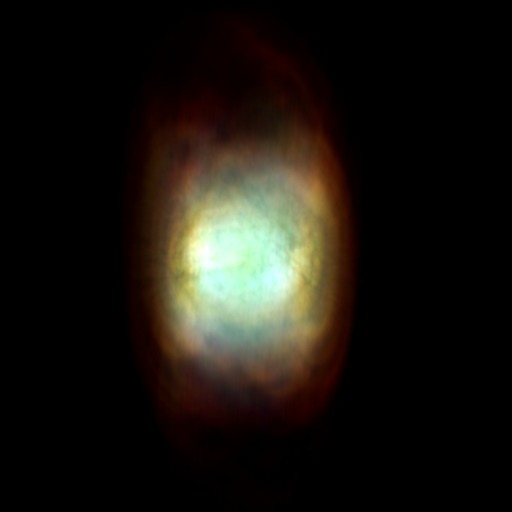}
\label{fig:ic4406}
}
\caption{Planetary nebulae NGC~6826, NGC~3132, and IC~4406.
From left to right: observed image and reconstruction using axial symmetry assumption seen from Earth and from space, respectively.
\emph{Original images: B. Balick (University of Washington), J. Alexander (University of Washington),
A. Hajian (U.S. Naval Observatory), Y. Terzian (Cornell University), M. Perinotto (University of Florence, Italy),
P. Patriarchi (Arcetri Observatory, Italy) and NASA;
The Hubble Heritage Team (STScI/AURA/NASA);
NASA and The Hubble Heritage Team (STScI/AURA)}}
\label{fig:eurovis-axial}\end{figure*}

Aside from a model of light emission and transport,
automatic reconstruction algorithms rely on assumptions about the geometry of the astronomical phenomenon we want to reconstruct.
Often, such assumptions can be deduced from the physical theories describing its formation and evolution.
For example, planetary nebulae and supernova remnants develop when a gaseous cloud is ejected from a dying star.
Since the star itself is approximately symmetric,
the cloud will exhibit some degree of symmetry unless external influences deform it.
Deformations may arise for many reasons,
including an irregular ambient medium or relative movement of the nebula with respect to the interstellar medium.
For single stars with negligible rotation and magnetic field,
the resulting nebulae are often of overall spherical symmetry, Figure~\ref{fig:eurovis-spherical}.
Otherwise, for example when fast rotation or a binary star are involved,
axial symmetries can occur, Figure~\ref{fig:eurovis-axial}.

Such symmetries can help resolve the depth ambiguity.
When an object exhibits only emission---no absorption or scattering---and is exactly axisymmetric or even spherically symmetric,
its three-dimensional geometry is uniquely defined by a single image.
A similar statement holds when the object absorbs light from some known source in the background
without emission or scattering within the object.
In both cases, reconstructing a symmetrical volume from a single image
amounts to solving a large system of linear equations~\cite{wenger2009algebraic},
applying a deconvolution method~\cite{leahy1991deprojection}
or optimizing the volumetric distribution subject to given constraints~\cite{magnor2005reconstruction}.
As soon as emission and absorption occur within the same object, such as in a star formation region, the ambiguity reemerges:
for example, light could be emitted and absorbed in the object without ever reaching the observer.
Scattering complicates this process even further.
Fortunately, for planetary nebulae, scattering and absorption can often be neglected.

While the assumption of exact symmetry completely resolves the depth ambiguity,
in reality even planetary nebulae are seldom exactly symmetric.
This may be due to interaction with the surrounding medium,
asymmetries in the initial conditions
or large-scale instabilities during the expansion of the gas.
Reconstruction algorithms, even those based on symmetry assumptions, have to take this into account:
a perfectly symmetric reconstruction from an only approximately symmetric image
will not be able to explain the observational data completely,
and will look unrealistic and rather boring in animation.
However, any part of the object that is in conflict with the symmetry assumption
raises the problem of missing depth information again,
and those parts have to be dealt with using heuristic algorithms~\cite{wenger2009algebraic}.
In this way, recent automatic approaches are able to create plausible high-resolution volumes rich in visual detail,
Figures~\ref{fig:eurovis-spherical} and \ref{fig:eurovis-axial},
albeit with less plausible geometry than manual approaches.

As we have seen before, the light transport in most planetary nebulae is particularly simple.
Reflection nebulae, which contain both gas and dust and therefore combine emission with absorption and scattering, are much more complex.
However, at radio wavelengths, the dust is almost transparent.
Thus, radio maps can be used to determine the emissive gas distribution
in conjunction with a symmetry assumption~\cite{lintu2007multi}
in the same way as for planetary nebulae.
Similarly, the dust distribution can be reconstructed
based on its thermal radiation in the infrared part of the spectrum~\cite{lintu2007emission}.
In both cases, the remaining distribution---gas or dust, respectively---is then uniquely defined by the optical image.

\begin{figure*}[ftbh]\centering
\subfigure{
\includegraphics[height=.1\linewidth]{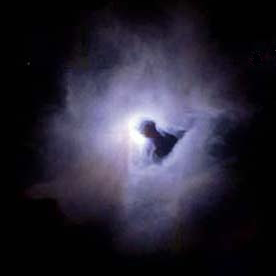}
\includegraphics[height=.1\linewidth]{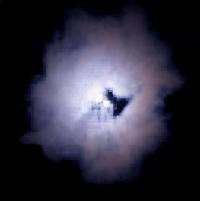}
\includegraphics[height=.1\linewidth]{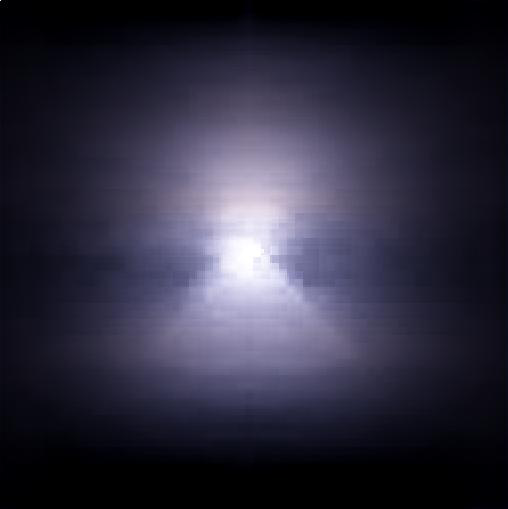}
\label{fig:ngc1999}
}
\subfigure{
\includegraphics[height=.1\linewidth]{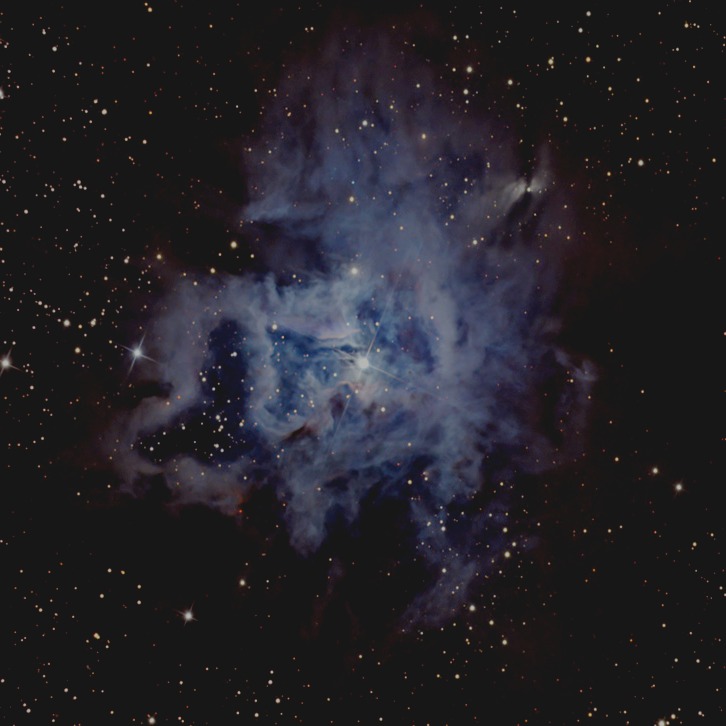}
\includegraphics[height=.1\linewidth]{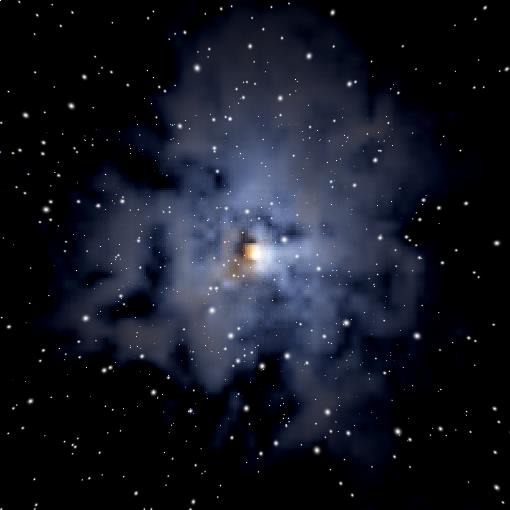}\label{fig:IrisNebula-front}
\includegraphics[height=.1\linewidth]{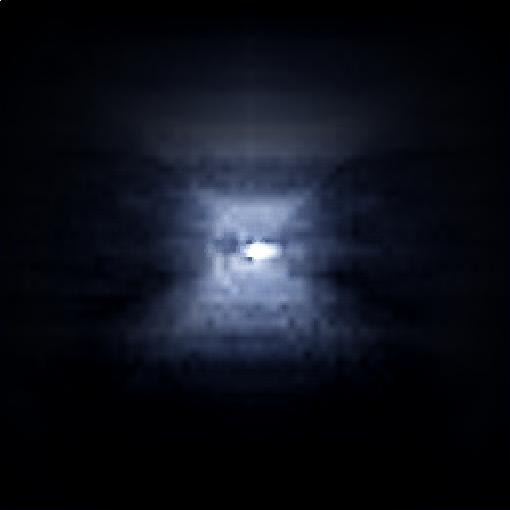}\label{fig:IrisNebula-oblique}
\label{fig:iris}
}
\subfigure{
\includegraphics[height=.1\linewidth]{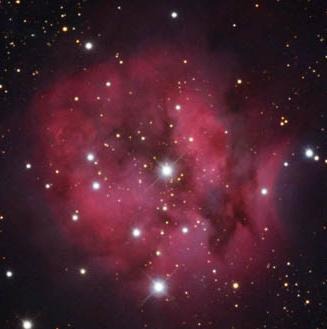}
\includegraphics[height=.1\linewidth]{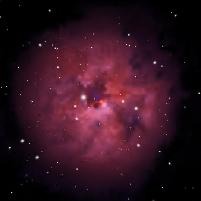}\label{fig:CocoonNebula-front}
\includegraphics[height=.1\linewidth]{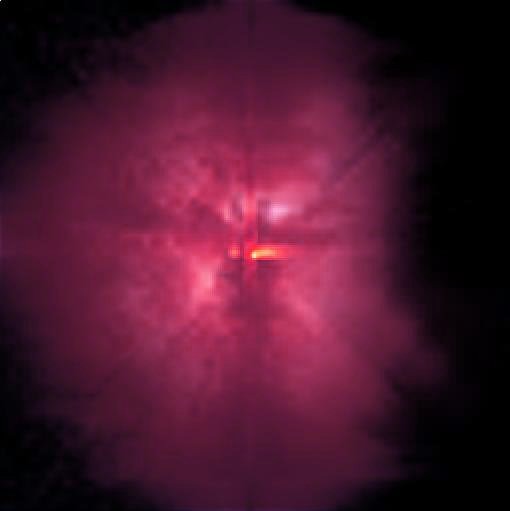}\label{fig:CocoonNebula-oblique}
\label{fig:cocoon}
}
\caption{Reflection nebulae NGC~1999, Iris Nebula, and Cocoon Nebula.
From left to right: observed image and analysis-by-synthesis reconstruction seen from Earth and from space, respectively.
\emph{Images: \cite{lintu2007reflection}}}
\label{fig:lintu}\end{figure*}

The gas and dust distributions of a reflection nebula can also be reconstructed simultaneously.
However, due to the complex mathematics of emission, absorption and scattering,
the problem cannot directly be formulated as solving a system of linear equations.
Instead, an \emph{analysis-by-synthesis} approach is used:
starting from a very simple initial guess for the gas and dust distributions,
the algorithm renders the model and compares it to the image.
The model is then iteratively fitted to the observational data,
making small adjustments that are compatible with prior assumptions about the geometry.
For example, instead of using a symmetry assumption as for planetary nebulae,
we can assume that dust in reflection nebulae is located primarily close to the central star of the nebula.
Based on this assumption, an analysis-by-synthesis algorithm~\cite{lintu2007reflection}
can generate plausible visualizations even for irregular nebulae, Figure~\ref{fig:lintu},
albeit at low resolution due to the computational complexity of the analysis-by-synthesis approach.

Even though the simple prior assumptions about the structure of astronomical objects are based on physical reasoning,
they are only qualitatively valid for general classes of objects.
The particularities of each individual object invariably cause deviations from the idealized model.
Current automatic reconstruction results are therefore at best plausible, but never reliably physically correct.
More complex and specific physical models---for example, including simulation of hydrodynamics
or ionization in a nebula---could increase confidence in the accuracy of the results,
but their computational complexity has hitherto prohibited their use in automatic reconstruction algorithms.
No automatic system is yet on par with the expertise of an astronomer in resolving ambiguities in astronomical data.

\section*{3D Modeling}
\label{sec:modeling}
Only astronomers are able to judge what the real geometry
of a planetary nebula or a supernova remnant looks like,
and even among specialists, there is often disagreement:
the morphology of many objects is the subject of current research.
The model that astronomers and astrophysicists build of an object
has to simultaneously explain different kinds of observational data.
The most important source of information is imagery,
whether from the visible, the infrared or the ultraviolet part of the spectrum,
but other sources of data often provide more depth information.

For example, planetary nebulae continuously expand.
The movement of the gas towards or away from Earth causes a \emph{Doppler shift} of the wavelength of the emitted light,
see box on page~\pageref{box:doppler}.
Since the emission wavelengths of the different ions in a planetary nebula's gaseous shell are constant and exactly known,
spectra with high wavelength resolution provide a way to determine the velocity component towards the observer with high accuracy.
Interestingly, velocity often correlates with position.
This is easy to see when we assume that the nebula was created by a single, instantaneous explosion:
after a certain time, faster particles have traveled further from the center of the explosion than slower ones.
In this case, the shape of the nebula can be derived directly from the spectra~\cite{sabbadin2006structure}.
However, the eruption of a nebula is often a continuous process,
and magnetic fields, hydrodynamic effects or interactions with the surrounding medium can influence the velocity distribution.
Thus, in general, all data have to be interpreted in the context of a holistic model of the formation and evolution of the object.

A complete model of an astronomical nebula comprises at least the current spatial distribution of different ions and
their velocity distribution as a function of position.
Based on these data, a rendering algorithm is able to reproduce typical types of observational data,
such as optical images or Doppler shift spectra.
In a typical workflow, the astronomer will create an initial model of the nebula based on their prior assumptions about its geometry.
They will then try to reproduce the observational data from the model,
making adjustments to the model until it is consistent with all available data.

The process of 3D modeling is facilitated by interactive 3D modeling systems.
Traditionally, such systems are used by artists to generate various types of three-dimensional content
for movies, animations, or games,
and most provide integrated rendering engines to quickly check the results to expect.
However, conventional modeling tools do not cover all physical properties
that are needed for the accurate representation of astronomical objects,
such as the velocity distribution of particles and the spectral characteristics of the different ions.
Ordinary rendering engines also cannot reproduce all types of observational data obtained from scientific instruments,
such as the Doppler shift spectra that provide information about particle velocity.

\begin{figure*}[ftbh]\centering
\includegraphics[width=\linewidth]{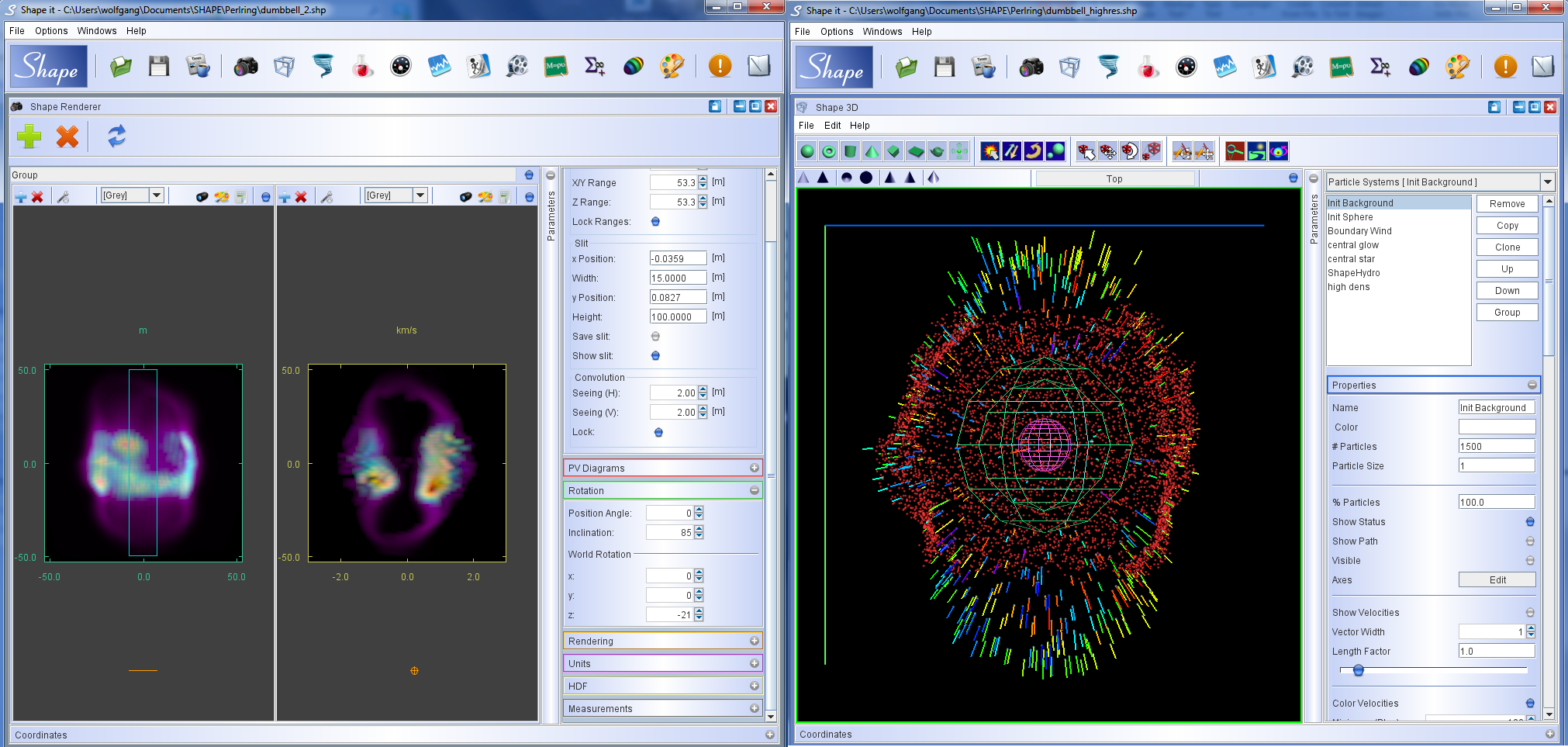}
\caption{Screenshots of the rendering (left) and 3D modeling (right) modules of the \emph{Shape} software.
The portrayed nebula was generated using \emph{Shape}'s hydrodynamical simulation capabilities.
The rendering module shows a false-color image (left) and a visualization of the Doppler shift of the spectral lines (right).
The screenshot of the 3D module illustrates the structure and velocity information obtained from the simulation.}
\label{fig:shape}\end{figure*}

Therefore, custom tools for astrophysical modeling and visualization have been developed.
The software \emph{Shape}~\cite{steffen2010shape} allows users
to interactively model astronomical objects and to generate physically accurate high-quality visualizations.
Scientists use it, among others, to validate their theories about formation and morphology of planetary nebulae.
Using the graphical interface of \emph{Shape}, Figure~\ref{fig:shape},
complex models can be built from simple deformable geometric primitives.
The models can be visualized and compared directly to observational data from within the software,
and selected parameters of the model can be automatically fitted to the data.

\section*{Space Hydrodynamics}
\label{sec:hydrodynamics}
Astronomers can only claim to understand the structure of an astronomical object
when they are able to explain its formation and evolution from appropriate initial conditions.
For planetary nebulae, such initial conditions might be
the characteristic physical variables of the dying progenitor star and its atmosphere.
In the simplest case, the structure of the emerging nebula will then be determined by the equations of hydrodynamics.
While many everyday fluid phenomena are readily described by the incompressible \emph{Navier-Stokes} equations,
this model is insufficient for the rarefied gases and high velocities that occur in astrophysics.
The compressible \emph{Euler} equations can be solved instead, see box on page~\pageref{box:hydrodynamics}.

\begin{figure*}[ftbh]\centering
\subfigure[]{\includegraphics[width=0.32\linewidth]{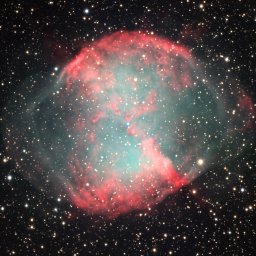}\label{fig:dumbbell-original}}
\subfigure[]{\includegraphics[width=0.32\linewidth]{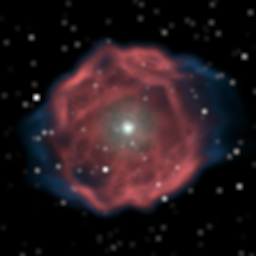}\label{fig:dumbbell-hydro}}
\subfigure[]{\includegraphics[width=0.32\linewidth]{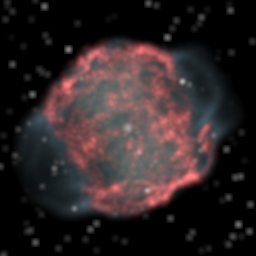}\label{fig:dumbbell-noise}}
\caption{The Dumbbell Nebula \subref{fig:dumbbell-original} and simulations done in \emph{Shape}.
The result of a hydrodynamic simulation is superimposed with manually modeled stars~\subref{fig:dumbbell-hydro}.
By adding a noise texture, the small-scale structure of the red high density gas can be modeled more accurately~\subref{fig:dumbbell-noise}.
\emph{Original image: Joe \& Gail Metcalf, Adam Block, NOAO, AURA, NSF}}
\label{fig:hydro}\end{figure*}

Most conventional 3D modeling tools provide some kind of fluid simulation for water or flame effects.
But because of the often supersonic motion in space, these simple hydrodynamic simulations are not
adequate for the simulation of astrophysical phenomena.
More complex simulation codes used in astrophysics research, on the other hand,
are usually designed by and for specialists.
Experimenting with such a simulation typically requires modifying the code, running the simulation,
and visualizing the results in a separate step.
\emph{Shape} provides a more intuitive front-end to astrophysical hydrodynamics simulations
by integrating the hydrodynamics simulation in the interactive graphical 3D modeling tool.
Simulation results can freely be combined with manually modeled parts, Figure~\ref{fig:hydro}.

By modeling 3D initial and boundary conditions directly in the graphical editor,
simulations can be run without having to do any programming.
Parameters and boundary conditions can be altered while the simulation is running and intermediate
results are visualized continuously.
This allows even artists and hydrodynamics novices to
quickly gain an intuition about astrophysical hydrodynamics.

\section*{Parallel GPU-Based Visualization}
\label{sec:rendering}
After an astronomical model has been created using automatic reconstruction, manual modeling, or simulation,
a method is needed to visualize it efficiently, accurately, and in high resolution.
Digital planetariums, for example, provide full dome multi-projector systems with high image resolutions
to create an inmersive experience for the audience.
Because creating high-quality images at the required resolution is a time-consuming task,
animation sequences are often pre-computed.
But pre-computed animations do not allow the presenters to interact with the spectators,
adapting their presentation while the show is running.
Therefore, modern planetarium systems like Digistar\textregistered~(\url{http://www.es.com})
or Uniview\texttrademark~(\url{http://www.scalingtheuniverse.com})
provide interfaces for visualizing digital content in real-time during a live performance.
While such interactive features offer great possibilities for novel presentations,
they also pose major challenges to the employed hardware and algorithms.

Many astronomical objects like nebulae or galaxies
can be conveniently represented by sampling the properties of the physical medium on a discrete three-dimensional grid. 
In scientific visualization, such data is commonly visualized using \emph{direct volume rendering}~\cite{callahan2008direct}.
For each pixel, a viewing ray is traced through the medium,
and the intensity is updated along the way according to the sampled values of light emission and absorption.
When scattering occurs, such as for reflection nebulae,
this simple scheme has to be extended
because light that reaches the camera can now originate from positions that the original viewing ray never reaches.
One solution is based on a multi-resolution rendering algorithm~\cite{magnor2005reflection}.
The volume is first sampled and rendered at very low resolution,
and each sampling point receives scattered intensity from the neighboring sampling points.
Because the resolution is low and each sampling points represents a large area within the volume,
this accounts for long-range scattering in the object.
The process it then repeated with successively increasing resolutions,
adding more and more detail to the rendered image.
Using this algorithm, reconstructed reflection nebula models can be rendered in real-time, Figure~\ref{fig:lintu}.

Most rendering algorithms require the casting of viewing rays through the volume to compute the color of the image pixels.
For interactive presentations, this step cannot be pre-computed,
and constitutes the most expensive operation especially for high-resolution images.
To achieve real-time performance, rendering can be accelerated
by performing many operations in parallel using graphics cards,
or \emph{graphics processing units} (GPUs)~\cite{engel2006realtime}.
However, large data sets easily exceed the memory capabilities of a single GPU.
High-performance real-time rendering algorithms for large data sets
therefore require the development of distributed parallel algorithms running on compute clusters containing multiple GPUs.

The key idea of distributed parallel volume rendering
is to decompose a large computational task into a set of smaller sub-tasks
that can be processed concurrently by a set of GPUs, or render nodes.
The two fundamental paradigms perform this decomposition either in image-space or in object-space~\cite{molnar1994sorting}.
The latter approach divides the volume into smaller chunks
that are distributed to the different nodes in the cluster and rendered separately.
Although this method scales easily to very large data sets,
combining the images from the different nodes for display
becomes increasingly costly when many nodes are involved.
However, the size of present-day astronomical models is still small enough
so that extreme data scalability is not the primary concern.
In contrast, the image resolution of digital productions is growing continuously---modern
projector systems feature resolutions of up to 8000$\times$8000 pixels---and
the increasing prevalence of stereoscopic shows requires to compute each image from two viewpoints simultaneously.

\begin{figure*}[ftbh]\centering
\subfigure[]{\includegraphics[height=.32\linewidth]{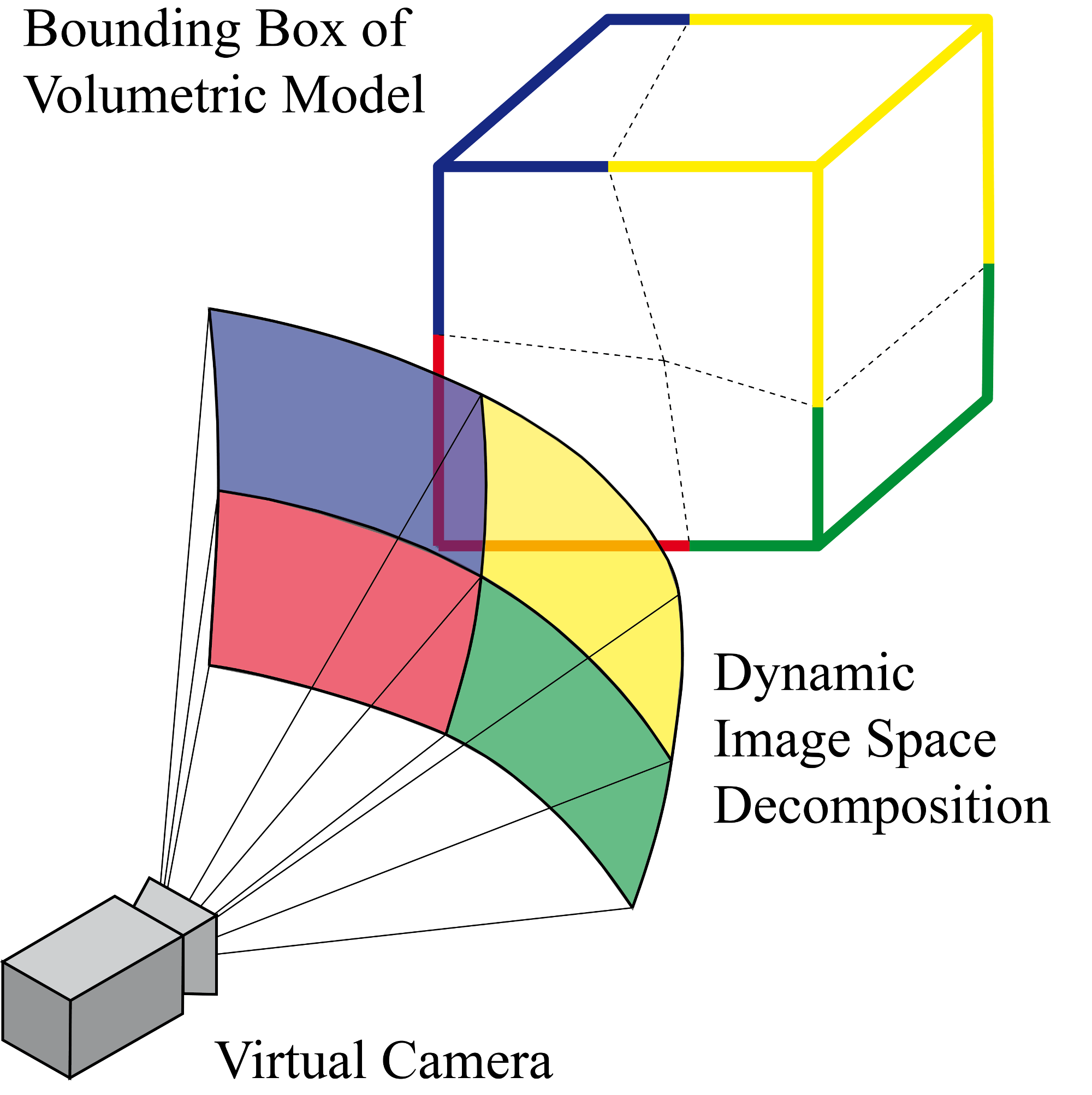}\label{fig:sortfist-schematic} }
\subfigure[]{\includegraphics[height=.32\linewidth]{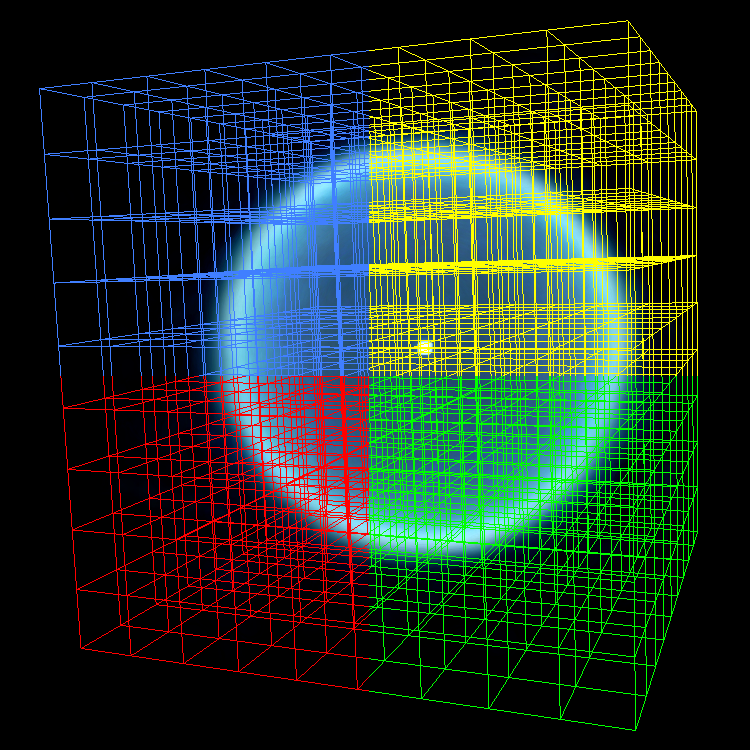}\label{fig:sortfirst-512bricks}}
\subfigure[]{\includegraphics[height=.32\linewidth]{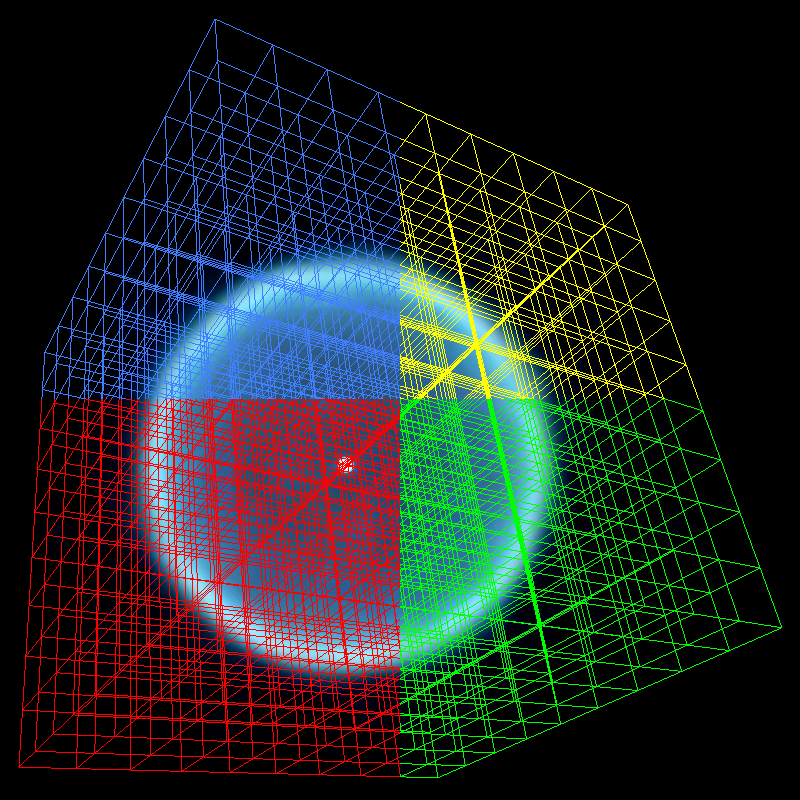}\label{fig:sortfirst-512bricks-2}}
\caption{In parallel volume rendering with dynamic image space decomposition,
the computation of different parts of the image is split up between the four render nodes,
here illustrated by different colors~\subref{fig:sortfist-schematic}.
To render its image tile, each node requires a different subset of the volumetric data,
here a model of the planetary nebula Abell~39~\subref{fig:sortfirst-512bricks}.
When the perspective changes, the required subsets of the data change accordingly~\subref{fig:sortfirst-512bricks-2},
and with adaptive image space decomposition, the size of the image tiles is recomputed such that each node receives the same workload.}
\label{fig:sortfirst}\end{figure*}

When the decomposition is performed in image-space,
each node computes a different region of the image.
Combined with multiple GPUs, image-space partitioning~\cite{moloney2011} is an efficient technique 
to compute images of very high resolution
because the number of pixels per node decreases linearly by adding more nodes to the cluster domain.
In contrast to an object-space decomposition, no additional compositing step is required,
but only a final gathering of the tiles for display.
Figure~\ref{fig:sortfirst} shows how the computational effort is distributed in a setup containing four render nodes.

Image-space decomposition in its simplest form requires
that the complete data set fits into the graphics memory of each compute node.
When this is not possible, the required amount of memory can be reduced
by dividing the data set into smaller bricks.
Each node then only has to store the bricks that are visible in its subsection of the image, Figure~\ref{fig:sortfirst-512bricks}.
A further performance improvement can be achieved by dynamically adapting the image space decomposition.
Whenever the viewpoint changes, the size of the image tiles is adapted such that each node has to perform the same amount of work.
Figure~\ref{fig:sortfirst-512bricks-2} shows an example of how the decomposition is adapted as the view point changes.

\section*{Outlook}
\label{sec:outlook}
The possibilities for astrophysical visualization and simulation have evolved enormously over the course of the last decades.
Still, the future holds many challenges in this field.
Computers are still not able to replace humans in the task of interpreting and understanding the world that surrounds us.
We have seen that also in astrophysical visualization,
manual work produces models that are far superior to automatic results
in terms of scientific accuracy and plausibility.
However, manual modeling of complex and detail-rich structures is extremely cumbersome.
In the future, we expect automatic reconstruction techniques to be integrated with existing manual modeling tools,
so that details are generated automatically while important geometrical features can be specified by an experienced astronomer.

In the area of astrophysical simulation, much improvement in terms of quality and speed
has been achieved by the development of better numerical algorithms and the constant increase in computing power of current processors.
Over the last years, however, this constant increase seems to have come to a halt;
instead, the number of processors in a single machine has started to increase.
Due to their inherent parallelism, hydrodynamics simulations are well suited for simultaneous computation on a large number of processors.
Notably, modern graphics cards provide the average customer with the technology
to compute even large and accurate simulations at interactive rates.
As this technology is being adopted in modeling and visualization tools,
it may revolutionize the way we explore and experience our cosmos.

\acknowledgments

This work was partially funded by Deutsche Forschungsgemeinschaft (DFG) as part of
the DFG project MA~2555/7-1 ``Astrographik''.
W.S. acknowledges support from {\em Universidad Nacional Aut\'onoma de M\'exico} - DGAPA and the Alexander von Humboldt Foundation.


\vspace{0.4\baselineskip}
\textbf{Stephan Wenger} is a researcher and PhD student at Braunschweig University of Technology, Braunschweig, Germany.
His research interests include reconstruction and visualization of astrophysical phenomena,
volume modeling and visualization techniques as well as computational photography.
He received Diplom (MSc) degrees in computer science and physics from Braunschweig University of Technology.
Contact him at \texttt{wenger@cg.cs.tu-bs.de}.

\vspace{0.4\baselineskip}
\textbf{Marco Ament} is a research assistant and PhD student at the University of Stuttgart, Stuttgart, Germany.
His research interests include direct volume rendering, parallel GPU methods, and flow visualization.
He received a Diplom (MSc) degree in computer science from the University of T{\"u}bingen, Germany.
Contact him at \texttt{marco.ament@visus.uni-stuttgart.de}.

\vspace{0.4\baselineskip}
\textbf{Wolfgang Steffen} is a professor at the {\em Universidad Nacional Aut\'onoma de M\'exico}, Ensenada, Mexico,
and visiting researcher at Braunschweig University of Technology, Germany.
His research interests include astrophysical hydrodynamics and computer graphics for astrophysics.
He received a PhD in astrophysics from the Max-Planck Institute for Radioastronomy and the University of Bonn, Germany.
Contact him at \texttt{wsteffen@astrosen.unam.mx}.

\vspace{0.4\baselineskip}
\textbf{Nico Koning} is a postdoctoral researcher at the University of Calgary, Calgary, Canada.
His research interests include astrophysics and the application of computer graphics methods to 
the modeling of astrophysical processes.
He received his PhD in astrophysics from the University of Calgary.
Contact him at \texttt{nico.koning@ucalgary.ca}.

\vspace{0.4\baselineskip}
\textbf{Daniel Weiskopf} is a professor at the University of Stuttgart, Stuttgart, Germany.
His research interests include visualization, visual analytics, GPU methods, computer graphics, and special and general relativity.
He received a PhD (Dr. rer. nat.) in physics from the University of T{\"u}bingen, Germany.
He is a member of IEEE Computer Society, ACM Siggraph, and the Gesellschaft f{\"u}r Informatik.
Contact him at \texttt{weiskopf@visus.uni-stuttgart.de}.

\vspace{0.4\baselineskip}
\textbf{Marcus Magnor} is a professor at Braunschweig University of Technology, Braunschweig, Germany.
His research interests meander along the visual information processing pipeline,
from image formation, acquisition, and analysis to image synthesis, display, perception, and cognition.
Ongoing research topics include image-based measuring and modeling, photo-realistic \& real-time rendering, and perception in graphics.
He received a PhD in electrical engineering from the University of Erlangen.
He is a member of IEEE Computer Society, ACM Siggraph, and the Gesellschaft f{\"u}r Informatik.
Contact him at \texttt{magnor@cg.cs.tu-bs.de}.

\bibliographystyle{abbrv}
\bibliography{cise}

\begin{thebibliography}{10}

\bibitem{callahan2008direct}
S.~P. Callahan, J.~H. Callahan, C.~E. Scheidegger, and C.~T. Silva.
\newblock Direct volume rendering: A {3D} plotting technique for scientific
  data.
\newblock {\em Computing in Science and Engineering}, 10(1):88--92, 2008.

\bibitem{engel2006realtime}
K.~Engel, M.~Hadwiger, J.~Kniss, C.~Rezk-Salama, and D.~Weiskopf.
\newblock {\em Real-Time Volume Graphics}.
\newblock A K Peters, 2006.

\bibitem{leahy1991deprojection}
D.~A. Leahy.
\newblock Deprojection of emission in axially symmetric transparent systems.
\newblock {\em Astronomy \& Astrophysics}, 247:584--589, 1991.

\bibitem{lintu2007reflection}
A.~Lin\c{t}u, L.~Hoffmann, M.~Magnor, H.~P.~A. Lensch, and H.-P. Seidel.
\newblock {3D} reconstruction of reflection nebulae from a single image.
\newblock In {\em Vision, Modeling, and Visualization}, pages 109--116, 2007.

\bibitem{lintu2007emission}
A.~Lin\c{t}u, H.~P.~A. Lensch, M.~Magnor, S.~El-Abed, and H.-P. Seidel.
\newblock {3D} reconstruction of emission and absorption in planetary nebulae.
\newblock In {\em IEEE/EG International Symposium on Volume Graphics}, pages
  9--16, 2007.

\bibitem{lintu2007multi}
A.~Lin\c{t}u, H.~P.~A. Lensch, M.~Magnor, T.-H. Lee, S.~El-Abed, and H.-P.
  Seidel.
\newblock A multi-wavelength-based method to de-project gas and dust
  distributions of several planetary nebulae.
\newblock In {\em Asymmetrical Planetary Nebulae IV}, pages 31--35, 2007.

\bibitem{magnor2005reflection}
M.~Magnor, K.~Hildebrand, A.~Lin\c{t}u, and A.~J. Hanson.
\newblock Reflection nebula visualization.
\newblock In {\em IEEE Visualization}, pages 255--262, 2005.

\bibitem{magnor2005reconstruction}
M.~Magnor, G.~Kindlmann, C.~Hansen, and N.~Duric.
\newblock Reconstruction and visualization of planetary nebulae.
\newblock {\em IEEE Transactions on Visualization and Computer Graphics},
  11(5):485--496, 2005.

\bibitem{magnor2010planetariums}
M.~Magnor, P.~Sen, J.~Kniss, E.~Angel, and S.~Wenger.
\newblock Progress in rendering and modeling for digital planetariums.
\newblock In {\em Eurographics Area Papers 2010}, pages 1--8, 2010.

\bibitem{molnar1994sorting}
S.~Molnar, M.~Cox, D.~Ellsworth, and H.~Fuchs.
\newblock A sorting classification of parallel rendering.
\newblock {\em IEEE Computer Graphics and Applications}, 14(4):23--32, jul
  1994.

\bibitem{moloney2011}
B.~Moloney, M.~Ament, D.~Weiskopf, and T.~M\"oller.
\newblock Sort first parallel volume rendering.
\newblock {\em IEEE Transactions on Visualization and Computer Graphics},
  17(8):1164--1177, 2011.

\bibitem{mueller2011general}
T.~M\"uller and D.~Weiskopf.
\newblock General-relativistic visualization.
\newblock {\em Computing in Science and Engineering}, 13(6):64--71, 2011.

\bibitem{mueller2011special}
T.~M\"uller and D.~Weiskopf.
\newblock Special-relativistic visualization.
\newblock {\em Computing in Science and Engineering}, 13(4):85--93, 2011.

\bibitem{raga2000adaptive}
A.~C. Raga, R.~Navarro-Gonz{\'a}lez, and M.~Villagr{\'a}n-Muniz.
\newblock A new, {3D} adaptive grid code for astrophysical and geophysical
  gasdynamics.
\newblock {\em Revista Mexicana de Astronom\'{\i}a y Astrof\'{\i}sica},
  36:67--76, 2000.

\bibitem{sabbadin2006structure}
F.~Sabbadin, M.~Turatto, R.~Ragazzoni, E.~Cappellaro, and S.~Benetti.
\newblock The structure of planetary nebulae: theory vs. practice.
\newblock {\em Astronomy \& Astrophysics}, 451:937--949, 2006.

\bibitem{steffen2010shape}
W.~Steffen, N.~Koning, S.~Wenger, C.~Morisset, and M.~Magnor.
\newblock Shape: {A} {3D} modeling tool for astrophysics.
\newblock {\em {IEEE} Transactions on Visualization and Computer Graphics},
  17(4):454--465, 2011.

\bibitem{wenger2009algebraic}
S.~Wenger, J.~Aja~Fern{\'{a}}ndez, C.~Morisset, and M.~Magnor.
\newblock Algebraic {3D} reconstruction of planetary nebulae.
\newblock {\em Journal of {WSCG}}, 17(1):33--40, 2009.

\end{thebibliography}

\end{document}